\documentclass{ptptex}

\usepackage[dvips]{graphicx}
\usepackage{latexsym}

\RequirePackage{hyperref}

\newcommand{\uD}{\ensuremath{\mathrm{d}}}
\newcommand{\uE}{\ensuremath{\mathrm{e}}}
\newcommand{\uI}{\ensuremath{\mathrm{i}}}

\renewcommand{\vec}[1]{\ensuremath{\boldsymbol{\mathrm{#1}}}}
\newcommand{\Cdot}{\ensuremath{\boldsymbol{\cdot}}}

\newcommand{\BK}[3]{\ensuremath{\left\langle{#1}\!%
\mathrel{\left|\vphantom{{#1}}{#2}\vphantom{{#3}}\right|%
\kern-\nulldelimiterspace}\!{#3}\right\rangle}}

\newcommand{\commute}[2]{\ensuremath{\left[{#1}\!%
\mathrel{\vphantom{{#1}},\vphantom{{#2}}%
\kern-\nulldelimiterspace}\!{#2}\right]}}
\newcommand{\anticommute}[2]{\ensuremath{\left\{{#1}\!%
\mathrel{\vphantom{{#1}},\vphantom{{#2}}%
\kern-\nulldelimiterspace}\!{#2}\right\}}}

\newcommand{\ArXivNo}{\href{http://arxiv.org/abs/quant-ph/0412105}{quant-ph/0412105}}

\newcommand{\XXXSize}{{\fontsize{12}{12}\selectfont\fbox{\textbf{\ArXivNo}}}}
\newcommand{\XXXTitle}{\hfill\XXXSize\newline\vskip 0.4cm}

\title{\XXXTitle{}A Derivation of the $Z\to\infty$ Limit for Atoms\thanks{%
Published in \textit{Progress of Theoretical Physics},
Vol.~\textbf{103}, No.~4, April 2000, pp.~697--702. \newline %
[\url{http://ptp.ipap.jp/link?PTP/103/697/}] %
}}

\author{\textsc{Edouard~B.~Manoukian}\thanks{E-mail: \texttt{edouard@sut.ac.th}}
and \textsc{Jarin~Osaklung}}

\inst{School of Physics, \ Suranaree University of Technology \\
Nakhon Ratchasima, 30000, Thailand}

\abst{Upper and lower bounds are derived for the ground-state
energy of neutral atoms which for $Z\to\infty$ both involve the
limits of exact Green's functions with one-body potentials. 
The limits of both bounds are shown to coincide with the
Thomas--Fermi ground-state energy.}

\begin{document}

\maketitle

\section{Introduction}
A very remarkable property of atoms is that in the limit
$Z\to\infty$, the Thomas--Fermi energy\cite{Thomas_1927,
Fermi_1927, Fermi_1928} becomes exact.\cite{Lieb_1976,
Lieb_1981, Baumgartner_1976} \   Unfortunately, the very ingenious
proofs of this beautiful result are somewhat complex.    We
have strived in developing a relatively easier, but rather formal,
derivation of this fundamental result for neutral atoms by using,
in the process, the Green's function corresponding to the 
Thomas--Fermi potential.    The derivation rests on the fact that
elementary scaling properties of integrals of the Green's function
allow one readily to consider the $Z\to\infty$ limit with no
difficulty.    The basic idea is that integrals of the Green's
function for coincident space points involved in the analysis have
particularly simple power law behaviour for large $Z$.    This
is spelled out in the text. 

For the Hamiltonian of neutral atoms we choose
\begin{equation}\label{Eqn01}
  H = \sum\limits_{\alpha=1}^{Z}\left(\frac{\vec{p}_{\alpha}^{2}}{2m}
  -\frac{Ze^{2}}{r_{\alpha}}\right)+\sum\limits_{\alpha<\beta}^{Z}
  \frac{e^{2}}{\left|\vec{r}_{\alpha}-\vec{r}_{\beta}\big.\right|}.
\end{equation}
We derive upper and lower bounds on the exact ground-state energy
of (\ref{Eqn01}), which for $Z\to\infty$ are the limits of
expressions involving integrals of the exact Green's functions
with one-body potentials.     The limits of both bounds are
shown to coincide with the ground-state Thomas--Fermi energy, thus
establishing the result.

\section{The upper bound}
We consider first the seemingly unrelated problem of a one-body
potential with Hamiltonian
\begin{equation}\label{Eqn02}
  h = \frac{\vec{p}^{2}}{2m}+V(\vec{r})
\end{equation}
where $V(\vec{r})$ is the Thomas--Fermi potential
\begin{align}
  V(\vec{r}) &= -\frac{Ze^{2}}{r}+e^{2}\int\!\uD^{3}\vec{r}'\:
  \frac{n(\vec{r}')}{\left|\vec{r}-\vec{r}'\big.\right|} \nonumber \\
  &= Z^{4/3}v(\vec{R})  \nonumber \\
  &\equiv{} -\frac{\hbar^{2}}{2m}\left(3\pi^{2}\big.\right)^{2/3}
  Z^{4/3}\left(\rho_{\mathrm{TF}}(\vec{R})\big.\right)^{2/3},
  \qquad{} \vec{r}=\frac{\vec{R}}{Z^{1/3}},
  \label{Eqn03}
\end{align}
and $n(\vec{r})=Z^{2}\rho_{\mathrm{TF}}(\vec{R})$ is the
Thomas--Fermi density normalized as
\begin{equation}\label{Eqn04}
  \int\!\uD^{3}\vec{r}\:n(\vec{r}) = Z.
\end{equation}
The Green's function corresponding to (\ref{Eqn02}) satisfies the
equation
\begin{equation}\label{Eqn05}
  \left[-\uI\frac{\partial}{\partial\tau}-\frac{\hbar^{2}}{2m}\nabla^{2}
  +V(\vec{r})\right]G_{\pm}(\vec{r}t,\vec{r}'0) = \delta^{3}(\vec{r}-\vec{r}')
  \delta(t),
\end{equation}
where, with appropriate boundary conditions,
\begin{equation}\label{Eqn06}
  G_{\pm}(\vec{r}t,\vec{r}'0) = \mp\left(\frac{\uI}{\hbar}\right)
  \mathrm{\Theta}(\mp{}t)\:G_{0}(\vec{r}\tau,\vec{r}'0;V), \qquad{}
  \tau=\frac{t}{\hbar}.
\end{equation}
We write
\begin{equation}\label{Eqn07}
  G_{0}(\vec{r}\tau,\vec{r}'0;V) = \int\!\!\frac{\uD^{3}\vec{k}}{(2\pi)^{3}}\;
  \uE^{\uI\vec{k}\Cdot(\vec{r}-\vec{r}')}\exp\left[-\uI\left(
  \frac{\hbar^{2}\vec{k}^{2}}{2m}\tau+U(\vec{r},\tau,\vec{k})\right)\right].
\end{equation}
We readily see that $U$ satisfies the equation
\begin{equation}\label{Eqn08}
  -\frac{\partial{}U}{\partial\tau}+V-\frac{\hbar^{2}}{m}\vec{k}\Cdot\vec{\nabla}U
  +\frac{\hbar^{2}}{2m}\left(\vec{\nabla}U\big.\right)^{2}+\uI\frac{\hbar^{2}}{2m}
  \nabla^{2}U = 0,
\end{equation}
with the boundary condition $U\big|_{\tau=0}=0$.     We are
particularly interested in the integral
\begin{equation}\label{Eqn09}
  \int\!\uD^{3}\vec{r}\;G_{0}(\vec{r}\tau,\vec{r}0;V),
\end{equation}
where $\exp\left[\uI\vec{k}\Cdot(\vec{r}-\vec{r}')\big.\right]$ in
(\ref{Eqn07}) becomes simply replaced by 1.     Under a scaling
we have $\vec{r}=\vec{R}/Z^{1/3}$, $V(\vec{r})=Z^{4/3}v(\vec{R})$,
where $v(\vec{R})$ is independent of $Z$.     Accordingly, to
study the large $Z$ behaviour, we carry out the change of
variables $\vec{r}\to\vec{R}$ and simultaneously substitute
$\tau=T/Z^{4/3}$.     Also, with the change of variables
$\vec{k}\to\vec{K}$, $\vec{k}=Z^{2/3}\vec{K}$, the product
$\vec{k}^{2}\tau=\vec{K}^{2}T$ in (\ref{Eqn07}) remains
invariant.  With these new variables, (\ref{Eqn08}) becomes
\begin{equation}\label{Eqn10}
  -\frac{\partial{}U}{\partial{}T}+v-\frac{\hbar^{2}}{mZ^{1/3}}
  \vec{K}\Cdot\vec{\nabla}_{\!\!R}U+\frac{\hbar^{2}}{2mZ^{2/3}}
  \left(\vec{\nabla}_{\!\!R}U\big.\right)^{2}+\uI\frac{\hbar^{2}}{2mZ^{2/3}}
  \nabla_{\!\!R}^{2}U = 0.
\end{equation}
Let $\lim_{Z\to\infty}U=U_{\infty}$.     Then (\ref{Eqn10})
collapses to $-\partial{}U_{\infty}/\partial{}t+v=0$, whose
solution is $U_{\infty}=vT$.     Hence for $Z\to\infty$, the
expression in (\ref{Eqn09}) becomes simply scaled by
$Z^{-1}Z^{2}=Z$.     [On the other hand, if we carry out the
unitary scale transformation $\vec{k}\to\vec{K}$,
$\vec{k}=Z^{1/3}\vec{K}$, (\ref{Eqn08}) leads to
$U=vT+\mathcal{O}\!\left(Z^{-2/3}\big.\right)$, and
$\left[\hbar^{2}\vec{k}^{2}\tau/2m+U\big.\right]\to
\left[\hbar^{2}\vec{K}^{2}T/2mZ^{2/3}+vT+\mathcal{O}\!\left(Z^{-2/3}\big.\right)
\right]$.     The latter, under the subsequent change of
variables $\vec{K}\to{}Z^{1/3}\vec{K}$, leads to
$\left[\hbar^{2}\vec{K}^{2}T/2m+vT+\mathcal{O}\!\left(Z^{-1/3}\big.\right)
\right]$, giving the same expression for (\ref{Eqn09}) as before,
with an overall scaling by $Z$.] 

Accordingly, we have the following limits for large $Z$, as
readily verified upon substitution of $vT$ for $U$, $Z\to\infty$~:
\begin{subequations}
\begin{equation}\label{Eqn11a}
  \int\!\uD^{3}\vec{r}\:\frac{2}{2\pi\uI}\int_{-\infty}^{\infty}\!
  \frac{\uD\tau}{\tau-\uI\varepsilon}\:G_{0}(\vec{r}\tau,\vec{r}0;V)
  \longrightarrow{}
  Z\int\!\uD^{3}\vec{R}\:\rho_{\mathrm{TF}}(\vec{R})\equiv{}Z,
\end{equation}
\begin{align}
  Z^{-7/3} & \int\!\uD^{3}\vec{r}\:\frac{2}{2\pi\uI}\int_{-\infty}^{\infty}\!
  \frac{\uD\tau}{\tau-\uI\varepsilon}\:\uI\frac{\partial}{\partial\tau}
  G_{0}(\vec{r}\tau,\vec{r}0;V) \nonumber \\
  \longrightarrow{}&
  2\int\!\uD^{3}\vec{R}\int\!\!\frac{\uD^{3}\vec{K}}{(2\pi)^{3}}\left[
  \frac{\hbar^{2}\vec{K}^{2}}{2m}+v(\vec{R})\right]\mathrm{\Theta}\!\left(
  \sqrt{-\frac{2mv(\vec{R})}{\hbar^{2}}}-|\vec{K}|\right) \nonumber \\
  &\quad{}=
  \left(3\pi^{2}\big.\right)^{5/3}\frac{\hbar^{2}}{10\pi^{2}m}\int\!\uD^{3}\vec{R}
  \:\left(\rho_{\mathrm{TF}}(\vec{R})\big.\right)^{5/3}-e^{2}\int\!\uD^{3}\vec{R}\:
  \frac{\rho_{\mathrm{TF}}(\vec{R})}{R} \nonumber \\
  &\quad\quad{}
  +e^{2}\int\!\uD^{3}\vec{R}\int\!\uD^{3}\vec{R}'\:\rho_{\mathrm{TF}}(\vec{R})
  \frac{1}{\left|\vec{R}-\vec{R}'\big.\right|}\rho_{\mathrm{TF}}(\vec{R}').
  \label{Eqn11b}
\end{align}
\end{subequations}
Here, the factor 2 multiplying the $\tau$-integrals is to account
for spin.     The $\tau$-integrals project out the negative
spectrum of $h$. 

Equation (\ref{Eqn11a}) in particular is of fundamental
importance.     It states that for large $Z$, the Hamiltonian
$h$, allowing for spin, has $Z$ (orthonormal) eigenvectors
corresponding to its negative spectrum.     Let
$g_{1}(\vec{r},\sigma),\ldots,g_{Z}(\vec{r},\sigma)$ denote these
eigenvectors for large $Z$.     Define the determinantal
(anti-symmetric) function
\begin{equation}\label{Eqn12}
  \phi_{Z}(\vec{r}_{1}\sigma_{1},\ldots,\vec{r}_{Z}\sigma_{Z}) =
  \frac{1}{\sqrt{Z!}}\:\det\left[g_{\alpha}(\vec{r}_{\beta},\sigma_{\beta})
  \big.\right].
\end{equation} 

Since such an anti-symmetric function does not necessarily
coincide with the ground-state function of the Hamiltonian $H$ in
(\ref{Eqn01}) in question, the expectation value
$\BK{\phi_{Z}}{H}{\phi_{Z}\big.}$ with respect to $\phi_{Z}$ in
(\ref{Eqn12}) can only \emph{overestimate} the exact ground-state
energy $E_{Z}$ of $H$, or at best be equal to it. 

We rewrite the Hamiltonian in (\ref{Eqn01}) equivalently as
\begin{equation}\label{Eqn13}
  H = \sum\limits_{\alpha=1}^{Z}h_{\alpha}+\left(
  \sum\limits_{\alpha<\beta}^{Z}\frac{e^{2}}{\left|\vec{r}_{\alpha}
  -\vec{r}_{\beta}\big.\right|}-e^{2}\sum\limits_{\alpha=1}^{Z}
  \int\!\uD^{3}\vec{r}'\:\frac{n(\vec{r}')}{\left|\vec{r}_{\alpha}
  -\vec{r}'\big.\right|}\right),
\end{equation}
where $h_{\alpha}$ is defined in (\ref{Eqn02}) with variables
$\vec{r}_{\alpha}$, $\vec{p}_{\alpha}$. 

Accordingly,
\begin{align}
  \lim\limits_{Z\to\infty}Z^{-7/3}E_{Z} &\leqslant{} \lim\limits_{Z\to\infty}Z^{-7/3}
  \BK{\phi_{Z}}{H}{\phi_{Z}\big.} \nonumber \\
  &= \lim\limits_{Z\to\infty}Z^{-7/3}\sum\limits_{\alpha=1}^{Z}
  \BK{g_{\alpha}}{h_{\alpha}}{g_{\alpha}\big.}
  +\lim\limits_{Z\to\infty}Z^{-7/3}F_{Z},
  \label{Eqn14}
\end{align}
where
\begin{align}
  F_{Z} =& -e^{2}\sum\limits_{\sigma}\int\!
  \frac{\uD^{3}\vec{r}\,\uD^{3}\vec{r}'}{\left|\vec{r}-\vec{r}'\big.\right|}\:
  n_{Z}(\vec{r}\sigma,\vec{r}\sigma)\,n(\vec{r}') \nonumber \\
  &\quad{}
  +\frac{e^{2}}{2}\sum\limits_{\sigma,\sigma'}\int\!
  \frac{\uD^{3}\vec{r}\,\uD^{3}\vec{r}'}{\left|\vec{r}-\vec{r}'\big.\right|}
  \left[n_{Z}(\vec{r}\sigma,\vec{r}\sigma)\,n_{Z}(\vec{r}'\sigma',\vec{r}'\sigma')
  -\left|n_{Z}(\vec{r}\sigma,\vec{r}'\sigma')\big.\right|^{2}\Big.\right],
  \label{Eqn15}
\end{align}
\begin{equation}\label{Eqn16}
  n_{Z}(\vec{r}\sigma,\vec{r}'\sigma') = \sum\limits_{\alpha=1}^{Z}
  g_{\alpha}(\vec{r},\sigma)\,g_{\alpha}^{*}(\vec{r}',\sigma'),
\end{equation}
or
\begin{align}
  F_{Z} \leqslant{} -e^{2}\int\!
  \frac{\uD^{3}\vec{r}\,\uD^{3}\vec{r}'}{\left|\vec{r}-\vec{r}'\big.\right|}
  &{}
  \left[n(\vec{r}')\left(\sum\limits_{\sigma}n_{Z}(\vec{r}\sigma,\vec{r}\sigma)\right)\right.
  \nonumber \\
  &\quad{}
  -\frac{1}{2}\left.\left(\sum\limits_{\sigma}n_{Z}(\vec{r}\sigma,\vec{r}\sigma)\right)
  \left(\sum\limits_{\sigma'}n_{Z}(\vec{r}'\sigma',\vec{r}'\sigma')\right)\right].
  \label{Eqn17}
\end{align}
However, we also have
\begin{align}
  \lim\limits_{Z\to\infty}Z^{-2}\sum\limits_{\sigma}n_{Z}(\vec{r}\sigma,\vec{r}\sigma)
  &=
  \lim\limits_{Z\to\infty}Z^{-2}\:\frac{2}{2\pi\uI}\int_{-\infty}^{\infty}\!
  \frac{\uD\tau}{\tau-\uI\varepsilon}\;G_{0}(\vec{r}\tau,\vec{r}0;V)
  \nonumber \\
  &\equiv{} \rho_{\mathrm{TF}}(\vec{R}),
  \label{Eqn18}
\end{align}
\begin{align}
  \lim\limits_{Z\to\infty}Z^{-7/3}\sum\limits_{\alpha=1}^{Z}
  \BK{g_{\alpha}}{h_{\alpha}}{g_{\alpha}\big.}
  &= \lim\limits_{Z\to\infty}\left(Z^{-7/3}\:2\sum\limits_{\lambda<0}\lambda\right)
  \nonumber \\
  &= \lim\limits_{Z\to\infty}Z^{-7/3}\int\!\uD^{3}\vec{r}\:\frac{2}{2\pi\uI}
  \int_{-\infty}^{\infty}\!\frac{\uD\tau}{\tau-\uI\varepsilon}
  \nonumber \\
  &\qquad\qquad\qquad\qquad\qquad{}\times
  \uI\frac{\partial}{\partial\tau}G_{0}(\vec{r}\tau,\vec{r}0;V),
  \label{Eqn19}
\end{align}
where $\sum_{\lambda<0}\lambda$ in $2\sum_{\lambda<0}\lambda$ is a
sum over all the negative eigenvalues of $h$ in (\ref{Eqn02}),
allowing for multiplicity but not spin degeneracy.     The
factor 2 takes the latter into account. 

From (\ref{Eqn14})--(\ref{Eqn19}) and (\ref{Eqn11b}), we finally
have
\begin{align}
  \lim\limits_{Z\to\infty}Z^{-7/3}E_{Z} &\leqslant{}
  \frac{\left(3\pi^{2}\big.\right)^{5/3}\hbar^{2}}{10\pi^{2}m}\int\!\uD^{3}\vec{R}\:
  \left(\rho_{\mathrm{TF}}(\vec{R})\big.\right)^{5/3}-e^{2}\int\!\uD^{3}\vec{R}\:
  \frac{\rho_{\mathrm{TF}}(\vec{R})}{R}  \nonumber \\
  &\qquad{}
  +\frac{e^{2}}{2}\int\!\uD^{3}\vec{R}\,\uD^{3}\vec{R}'\:\rho_{\mathrm{TF}}(\vec{R})
  \frac{1}{\left|\vec{R}-\vec{R}'\big.\right|}\rho_{\mathrm{TF}}(\vec{R}'),
  \label{Eqn20}
\end{align}
and the right-hand side is the coefficient of $Z^{7/3}$ of the
ground-state Thomas--Fermi energy.

\section{The lower bound}
Given any arbitrary real and positive function
$\rho_{Z}(\vec{r})$, we use the following elementary text-book
bound\cite{Thirring_1981}~:
\begin{align}
  \sum\limits_{\alpha<\beta}^{Z}
  \frac{1}{\left|\vec{r}_{\alpha}-\vec{r}_{\beta}\big.\right|}
  \geqslant{}&
  \sum\limits_{\alpha=1}^{Z}\int\!\uD^{3}\vec{r}\:
  \frac{\rho_{Z}(\vec{r})}{\left|\vec{r}-\vec{r}_{\alpha}\big.\right|}
  -\frac{1}{2}\int\!\uD^{3}\vec{r}\,\uD^{3}\vec{r}'\:\rho_{Z}(\vec{r})
  \frac{1}{\left|\vec{r}-\vec{r}'\big.\right|}\rho_{Z}(\vec{r}')
  \nonumber \\
  &\quad{}
  -\frac{3}{2}\pi^{1/3}Z^{2/3}\left[\int\!\uD^{3}\vec{r}\;
  \left(\rho_{Z}(\vec{r})\big.\right)^{2}\right]^{1/3}.
  \label{Eqn21}
\end{align}
Here the real function $\rho_{Z}(\vec{r})$ may be chosen to be
positive and is otherwise \emph{arbitrary} (i.e., may be chosen at
will) to the extent that the integrals on the right-hand side of
(\ref{Eqn21}) exist.     We conveniently choose it in such a
way that $\rho_{Z}(\vec{r})\to{}Z^{2}\rho_{\mathrm{TF}}(\vec{R})$
for $Z\to\infty$, which will then coincide with $n(\vec{r})$ used
above in (\ref{Eqn03}).     Consider the Hamiltonian
$h'=\vec{p}^{2}/2m+V'$, where
\begin{equation}\label{Eqn22}
  V'(\vec{r}) = -\frac{Ze^{2}}{r}+e^{2}\int\!\uD^{3}\vec{r}'\:
  \frac{\rho_{Z}(\vec{r}')}{\left|\vec{r}-\vec{r}'\big.\right|}.
\end{equation}
With $\rho_{Z}(\vec{r})$ conveniently chosen, $V'(\vec{r})$ may be
chosen to be a locally square integrable function satisfying
$V'(\vec{r})\to{}0$ for $r\to\infty$.     Let $\psi$ be a
normalized antisymmetric function in
$(\vec{r}_{1}\sigma_{1},\ldots,\vec{r}_{Z}\sigma_{Z})$.   Then
(\ref{Eqn21}) implies that
\begin{align}
  \BK{\psi}{H}{\psi\big.}
  \geqslant{}&
  \BK{\psi}{\sum\limits_{\alpha}h'_{\alpha}}{\psi}
  -\frac{e^{2}}{2}\int\!\uD^{3}\vec{r}\,\uD^{3}\vec{r}'\:\rho_{Z}(\vec{r})
  \frac{1}{\left|\vec{r}-\vec{r}'\big.\right|}\rho_{Z}(\vec{r}')
  \nonumber \\
  &\quad{}
  -\frac{3}{2}\pi^{1/3}Z^{2/3}e^{2}\left[\int\!\uD^{3}\vec{r}\;
  \left(\rho_{Z}(\vec{r})\big.\right)^{2}\right]^{1/3}.
  \label{Eqn23}
\end{align} 

Consider the lowest energy $E$ of the Hamiltonian
$\sum_{\alpha}h'_{\alpha}$.     The Pauli exclusion principle
comes to the rescue here.\cite{Lieb_1976} \    Concerning the
Hamiltonian $\sum_{\alpha}h'_{\alpha}$, the $Z$
``non-interacting'' electrons (although each interacts with an
external potential $V'$) can be put, according to the Pauli
exclusion principle, in the lowest energy levels of
$\sum_{\alpha}h'_{\alpha}$ (allowing for spin degeneracy) if $Z$
is less than the number of such available levels.     If $Z$ is
larger, then the remaining free electrons should have arbitrarily
small kinetic energies to define the lowest energy of
$\sum_{\alpha}h'_{\alpha}$.     In either case,
$E\geqslant{}2\sum_{\lambda<0}\lambda$, where
$\sum_{\lambda<0}\lambda$, defined as above, is now applied to
$h'$.     Accordingly,
\begin{equation}\label{Eqn24}
  \lim\limits_{Z\to\infty}Z^{-7/3}\BK{\psi}{H}{\psi\big.}
  \geqslant{} \lim\limits_{Z\to\infty}K_{Z},
\end{equation}
where
\begin{align}
  K_{Z} &= Z^{-7/3}\int\!\uD^{3}\vec{r}\:\frac{2}{2\pi\uI}\int_{-\infty}^{\infty}\!
  \frac{\uD\tau}{\tau-\uI\varepsilon}\:\uI\frac{\partial}{\partial\tau}
  G_{0}(\vec{r}\tau,\vec{r}0;V')  \nonumber \\
  &\qquad{}
  -Z^{-7/3}\:\frac{e^{2}}{2}\int\!\uD^{3}\vec{r}\,\uD^{3}\vec{r}'\:\rho_{Z}(\vec{r})
  \frac{1}{\left|\vec{r}-\vec{r}'\big.\right|}\rho_{Z}(\vec{r}')
  \nonumber \\
  &\qquad{}
  -\frac{3}{2}\pi^{1/3}Z^{-5/3}e^{2}\left[\int\!\uD^{3}\vec{r}\;
  \left(\rho_{Z}(\vec{r})\big.\right)^{2}\right]^{1/3},
  \label{Eqn25}
\end{align}
and $G_{0}(\vec{r}\tau,\vec{r}0;V')$ is defined as above.   Also
here we have used the equality on the extreme right-hand side of
(\ref{Eqn19}).     Since the right-hand side of the inequality
(\ref{Eqn24}) is independent of $\psi$, this inequality holds with
$\psi$ corresponding to the ground-state function of $H$ as well,
i.e., with $\BK{\psi}{H}{\psi\big.}$ corresponding to
\begin{equation}\label{Eqn26}
  \min\limits_{\psi}\BK{\psi}{H}{\psi\big.} = E_{Z}.
\end{equation}
To the extent that $\rho_{Z}(\vec{r})>0$ is arbitrary, we choose
it conveniently as
\begin{equation}\label{Eqn27}
  \rho_{Z}(\vec{r}) =
  Z^{2}\rho_{\mathrm{TF}}(\vec{R})\sqrt{1-\uE^{-Z\alpha{}R}},
\end{equation}
where $\alpha>0$ is an arbitrary scale parameter.     We note
that $\rho_{\mathrm{TF}}(\vec{R})\sim{}R^{-3/2}$ for $R\to{}0$,
and that $\rho_{\mathrm{TF}}(\vec{R})\sim{}R^{-6}$ for
$R\to\infty$.     The factor $\sqrt{1-\exp(-Z\alpha{}R)}$
ensures the integrability of the last integral on the right-hand
side of (\ref{Eqn25}).     We estimate the latter for
$Z\to\infty$ as
\begin{align}
  \frac{1}{Z^{2/3}} & \left[\int\!\uD^{3}\vec{R}\;\rho_{\mathrm{TF}}^{2}(\vec{R})
  \left(1-\uE^{-Z\alpha{}R}\Big.\right)\right]^{1/3} \nonumber \\
  &\qquad\qquad\leqslant{}
  \left[\int_{\alpha{}R\leqslant{}1/Z}\!\uD^{3}\vec{R}\;\frac{\alpha}{Z}\:R\,
  \rho_{\mathrm{TF}}^{2}(\vec{R})\right. \nonumber \\
  &\qquad\qquad\qquad\quad{}
  +\frac{\left(1-\uE^{-Z}\Big.\right)}{Z^{2}}\int_{1>\alpha{}R>1/Z}\!\uD^{3}\vec{R}\;
  \rho_{\mathrm{TF}}^{2}(\vec{R})  \nonumber \\
  &\qquad\qquad\qquad\quad{}
  +\left.\frac{1}{Z^{2}}\int_{\alpha{}R\geqslant{}1}\!\uD^{3}\vec{R}\;
  \rho_{\mathrm{TF}}^{2}(\vec{R})\right]^{1/3}.
  \label{Eqn28}
\end{align}
The second integral on the right-hand side is at worst logarithmic
in $Z$.     Hence the last term on the right-hand side of
(\ref{Eqn25}) vanishes for $Z\to\infty$.     Since
${-\left(1-\uE^{-Z\alpha{}R}\big.\right)}\geqslant{}{-1}$, the
second term (with the minus sign) on the right-hand side of
(\ref{Eqn25}) is bounded below by
\begin{equation*}
  -\frac{e^{2}}{2}\int\!\uD^{3}\vec{R}\,\uD^{3}\vec{R}'\:\rho_{\mathrm{TF}}(\vec{R})
  \frac{1}{\left|\vec{R}-\vec{R}'\big.\right|}\rho_{\mathrm{TF}}(\vec{R}').
\end{equation*}
Finally, we note that since $1-\uE^{-Z\alpha{}R}\to{}1$ for
$Z\to\infty$, and with $V'\equiv{}Z^{4/3}v'_{Z}$ and
$\lim_{Z\to\infty}v'_{Z}\equiv{}v(\vec{R})$, the limit of the
first expression on the right-hand side of (\ref{Eqn25}) coincides
with that in (\ref{Eqn11b}) for $Z\to\infty$.     All told, we
see that the lower bound in (\ref{Eqn24}) coincides with the upper
bound in (\ref{Eqn20}).     This completes our
demonstration. 

In a future report, we will investigate to what extent this
analysis may be extended to other interactions.

\section*{Acknowledgements}
The authors would like to thank a referee for valuable
comments.

\end{document}